\documentclass[aps,preprint]{revtex4}
\usepackage[dvips]{graphicx}
\begin{document}
\draft
\title{\bf Experimental investigation of Wigner's reaction matrix for irregular graphs with absorption}

\author{Oleh Hul$^1$, Oleg Tymoshchuk$^1$, Szymon Bauch$^1$, Peter M. Koch$^2$, and Leszek Sirko$^1$}
\address{$^1$Institute of Physics, Polish Academy of Sciences,
Aleja \ Lotnik\'{o}w 32/46, 02-668 Warszawa, Poland \\
$^2$Department of Physics and Astronomy, State University of New
York, Stony Brook, NY 11794-3800 USA}

\date{September 22, 2005}

\bigskip

\begin{abstract}

We use tetrahedral microwave networks consisting of coaxial cables
and attenuators connected by $T$-joints to make an experimental
study of Wigner's reaction $K$ matrix for irregular graphs in the
presence of absorption. From measurements of the scattering matrix
$S$ for each realization of the microwave network we obtain
distributions of the imaginary and real parts of $K$. Our
experimental results are in good agreement with theoretical
predictions.

\end{abstract}


\bigskip
\maketitle

\smallskip

Pauling introduced quantum graphs of connected one-dimensional
wires almost seven decades ago \cite{Pauling}. Kuhn used the same
idea a decade later \cite{Kuhn} to describe organic molecules by
free electron models. Quantum graphs can be considered as
idealizations of physical networks in the limit where the lengths
of the wires greatly exceed their widths; this corresponds to
assuming that propagating waves remain in a single transverse
mode.  Among the systems modeled by quantum graphs are, e.g.,
electromagnetic optical waveguides \cite{Flesia,Mitra}, mesoscopic
systems \cite{Imry, Kowal} , quantum wires \cite{Ivchenko,
Sanchez} and excitation of fractons in fractal structures
\cite{Avishai, Nakayama}. Recent work has shown that quantum
graphs provide an excellent system for studies of quantum chaos
\cite{Kottossmilansky,Kottos,Prlkottos,Zyczkowski,Kus,Tanner,%
    Kottosphyse,Kottosphysa,Gaspard,Blumel,Hul2004}.
Quantum graphs with external leads (antennas) have been analyzed
in detail in \cite{Kottosphyse,Kottosphysa}. Quantum graphs with
absorption, a more realistic but more complicated system, have
been studied numerically in \cite{Hul2004}, but until now there
have been no experimental studies of the effect of absorption.

This paper presents results of our experimental study of
distributions  of Wigner's reaction matrix \cite{Akguc2001} (often
called in the literature just the $K$ matrix \cite{Fyodorov2004})
for microwave networks that correspond to graphs with time
reversal symmetry ($\beta=1$ symmetry class of random matrix
theory \cite{Mehta}) in the presence of absorption. For the case
of an experiment having a single-channel antenna, the $K$ matrix
and scattering matrix $S$ are related by
\begin{equation}
\label{Eq.1} S=\frac{1-iK}{1+iK}.
\end{equation}
The function $Z=iK$ has direct physical meaning as the electrical
impedance, which has been recently measured in a microwave cavity
experiment \cite{Anlage2005}. For the one-channel case the $S$
matrix can be parameterized as
\begin{equation}
\label{Eq.2} S=\sqrt{R}e^{i\theta},
\end{equation}
where $R$ is the reflection coefficient and $\theta$ is the phase.

After seminal work of L\'opez, Mello and Seligman \cite{Lopez1981}
came theoretical studies of the properties of statistical
distributions of the $S$ matrix with direct processes and
imperfect coupling \cite{Doron1992,Brouwer1995,Savin2001}. A
recent experiment investigated the distribution of the $S$ matrix
for chaotic microwave cavities with absorption \cite{Kuhl2005}.
The distribution $P(R)$ of the reflection coefficient $R$ in
Eq.~(2), at the beginning investigated in the strong absorption
limit \cite{Kogan}, has been recently known for any dimensionless
absorption strength $\gamma =2\pi \Gamma /\Delta$, where $\Gamma$
is the absorption width and $\Delta$ is the mean level spacing.
For systems with time reversal symmetry ($\beta=1$)
M\'endez-S\'anchez et al.\  \cite{Sanchez2003} studied $P(R)$
experimentally, and Savin et al.\   \cite{Savin2005} found an
exact formula for $P(R)$. For systems violating time reversal
symmetry ($\beta=2$), Beenakker and Brouwer \cite{Beenakker2001}
calculated $P(R)$ for the case of a perfectly coupled,
single-channel lead.

In our experiment we simulate quantum graphs with microwave
networks. The analogy between them is based on the Schr\"odinger
equation for the former being equivalent to the telegraph equation
for the latter \cite{Hul2004}. We call them microwave graphs.
Measurements of the scattering matrix for them were stimulated by
\cite{Blumel88} and the pioneering measurements in \cite{Doron90}.

A simple microwave graph, the tetrahedral case, consists of six
coaxial cables (bonds) that meet three-at-a-time at $N=4$
different $T$-joints (vertices). Each coaxial cable consists of an
inner conductor with radius $r_1$ separated from a concentric
outer conductor with inner radius $r_2$ by a homogeneous,
non-magnetic material with dielectric constant $\varepsilon$. The
fundamental $TEM$ mode that propagates (the so-called Lecher wave)
down to zero frequency exists because the cross section of the
cable is doubly connected \cite[p.\ 253]{Jones}. For frequencies
$\nu$ below the onset of the TE$_{11}$ mode in a coaxial cable,
which propagates above $\nu_{c} \simeq \frac{c}{\pi (r_1+r_2)
 \sqrt{\varepsilon}}$ \cite{Jones}, the cable is single mode: only the TEM mode propagates.
For SMA-RG-402 coaxial cable, which has $r_1 = 0.05$ cm, $r_2= 0.15$ cm,
and $\varepsilon \simeq 2.08$ (teflon dielectric), single-mode propagation
occurs below 32.9 GHz.

An (ideal) microwave graph with no aborption and no leads to the
outside world is a closed (bound) system.  The presence of
absorption and/or leads to the outside world creates an open
system. Because the coaxial cables are lossy, we may vary
absorption in the microwave graphs by changing the length of
cable(s), \cite{Hul2004}, by adding one or more (coaxial)
microwave attenuators, or by changing the coupling to the outside
world.

\begin{figure}[!]
\begin{center}
\rotatebox{270} {\includegraphics[width=0.5\textwidth,
height=0.6\textheight, keepaspectratio]{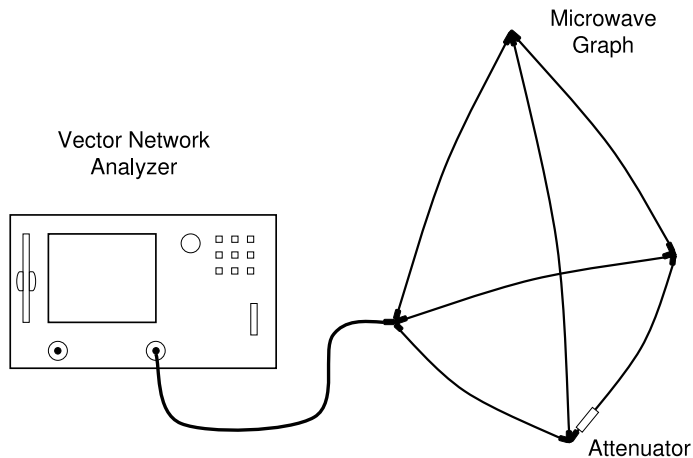}} \caption{A diagram
of experimental setup used to measure the scattering matrix $S$ of
tetrahedral microwave graphs with absorption. Absorption in the
graphs was varied by changing the attenuator. The vector network
analyzer used for all measurements was an HP model 8722D.
}\label{Fig1}
\end{center}
\end{figure}

Figure~1 shows our experimental setup for measurements of the
single-channel scattering matrix $S$ for tetrahedral microwave
graphs. We used a Hewlett-Packard model 8722D microwave vector
network analyzer to measure the scattering matrix $S$ of such
graphs in two different frequency windows, viz., 3.5--7.5 GHz and
12-16 GHz. As the figure shows, at one of the four vertices we
used a 4-joint connector to couple the microwave graph to the
vector network analyzer via a single-channel lead realized with an
HP model 85133-60017, low-loss, flexible microwave cable; the
other three vertices consisted of $T$-joints. The plane of
calibration in the measurements was at the entrance to the 4-joint
connector. Note that the microwave graph in Fig.~1 has a microwave
attenuator in one of its bonds.

To investigate the distributions of imaginary and real parts of
the $K$ matrix we measured the scattering matrix $S$ for $184$
different realizations of tetrahedral microwave graphs having a
microwave (SMA) attenuator in one of the bonds. For each graph
realization, which was obtained either by the replacement of the
bonds or putting an attenuator to a different bond, the scattering
matrix $S$ was measured in 1601 equally spaced steps. The total
optical lengths of the microwave graphs, including joints and the
single attenuator, was 196.2 cm when a 3 dB, 6 dB, or 20 dB
attenuator was used, whereas it was 197.4 cm when the 10 dB
attenuator was used. To avoid degeneracy of eigenvalues in the
graphs, we chose optical lengths for the bonds that were not
simply commensurable.

\begin{figure}[!]
\begin{center}
\rotatebox{0} {\includegraphics[width=0.5\textwidth,
height=0.8\textheight, keepaspectratio]{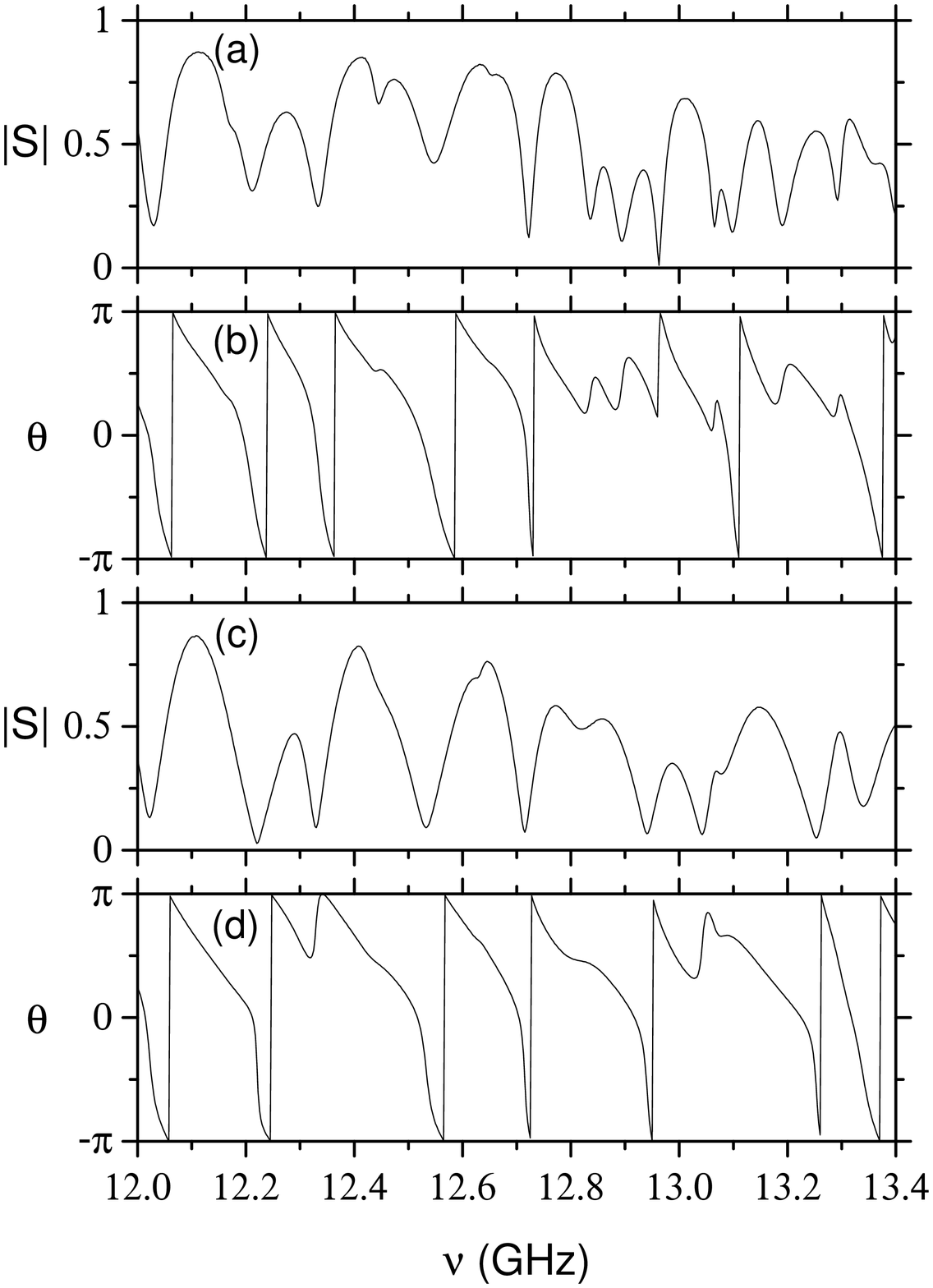}} \caption{Panels (a)
and (b) show, respectively, the modulus $|S|$ and the phase
$\theta $ of the scattering matrix $S$ measured for the graph (see
Fig.~1) with absorption parameter $\gamma = 3.6$ (see the text)
over the frequency range 12 - 13.4 GHz and with use of a 3 dB
attenuator. Panels (c) and (d) show corresponding measurements for
a graph with $\gamma = 6.8$ over the same frequency range and with
use of a 20 dB attenuator.  The total optical length, 196.2 cm, of
both microwave graphs was the same, including joints and the
attenuator.}\label{Fig2}
\end{center}
\end{figure}

Figure~2 shows the modulus $|S|$ and the phase $\theta $ of the
scattering matrix $S$ of a tetrahedral graph with $\gamma = 3.6$
(in panels (a) and (b), obtained with use of a 3 dB attenuator)
and one with $\gamma = 6.8$ (in panels (c) and (d), obtained with
use of a 20 dB attenuator); both cases cover the frequency range
12--13.4 GHz.  The lengths of corresponding bonds in the two
graphs were the same.  Direct processes are present in the
scattering because the microwave vector analyzer was connected to
the graphs by the 4-joint connector.  For each, individual
realization of the graph we may estimate them from the average
value of the scattering matrix $\langle S\rangle$. Our
measurements averaged over all realizations of microwave graphs
gave $\langle S\rangle_{av} \simeq 0.47\pm 0.03 + i(-0.01 \pm
0.04)$, where $i=\sqrt{-1}$. The experimental value for $|\langle
S\rangle_{av}| \simeq 0.47$ is close to a theoretical estimate
\cite{Kottosphyse,Kottosphysa} for the modulus of the vertex
reflection amplitude $|\rho|=0.5$ for a 4-joint connector with
Neumann boundary conditions,
\begin{equation}
\label{Eq.3} \rho=\frac{2}{n_v}-1,
\end{equation}
where $n_v=4$ is the number of bonds meeting at the vertex in
question.

Equation~(1) holds for systems with absorption but without direct
processes. The case of imperfect coupling $|\langle S\rangle| > 0$
and direct processes present can be mapped onto that of perfect
one \cite{Fyodorov2004} by making the following parametrization,
\begin{equation}
\label{Eq.4} S_0=\frac{S-|\langle S\rangle|}{1-|\langle S\rangle|
S},
\end{equation}
where $S_0$ is the scattering matrix of a graph for the
perfect-coupling case (no direct processes present).

For systems with time reversal symmetry ($\beta=1$ in equations
below), the distributions $P(v)$ of the imaginary and $P(u)$ of
the real parts of the $K$ matrix \cite{Fyodorov2004} are given by
the following interpolation formulas:
\begin{equation}
\label{Eq.5} P(v) = \frac{N_{\beta}e^{-a}}{\pi\sqrt{2\gamma }
v^{3/2}}(A[K_0(a)+K_1(a)]a+\sqrt{\pi}Be^{-a}),
\end{equation}
and
\begin{equation}
\label{Eq.6} P(u) = \frac{N_{\beta}e^{-\gamma /4}}{2\pi
\bar{u}}[\frac{A}{2}\sqrt{\frac{\gamma}{4}}D(\frac{\bar{u}}{2}) +
BK_1(\frac{\gamma \bar{u}}{4})],
\end{equation}
where $-v=\textrm{Im} \, K<0$ and $u=\textrm{Re} \,K$ are,
respectively, the imaginary and real parts of the $K$ matrix. The
normalization constant is $N_{\beta} =\alpha \left( A\Gamma(\beta
/2 +1,\alpha)+ Be^{-\alpha} \right)^{-1}$, where $\alpha=\gamma
\beta /2$, $\Gamma(x,\alpha)=\int_{\alpha}^{\infty}dt
t^{x-1}e^{-t}$ is the upper incomplete Gamma function,
$A=e^{\alpha} -1 $ and $B=1+\alpha -e^{\alpha}$. In Eq.~(5) the
variable $a=\frac{\gamma }{16}(\sqrt{v}+1/\sqrt{v})^2$ and  $K_0$,
$K_1$ are MacDonald functions. In Eq.~(6)
$D(z)=\int_0^{\infty}dq\sqrt{1+z(q+q^{-1})}e^{-\gamma
z(q+q^{-1})/4}$ and $\bar{u}=\sqrt{u^2+1}$.

\begin{figure}[!]
\begin{center}
\rotatebox{270} {\includegraphics[width=0.5\textwidth,
height=0.6\textheight, keepaspectratio]{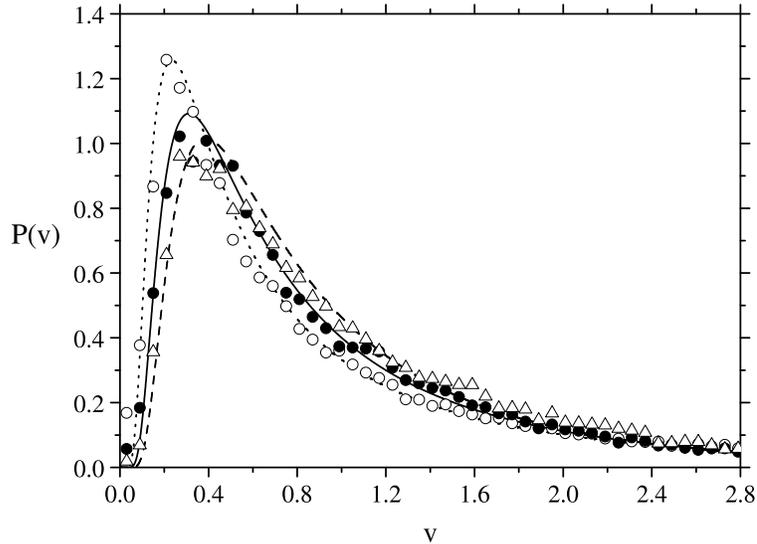}}
\caption{Experimental distribution $P(v)$ of the imaginary part of
the $K$ matrix  at different values of the mean absorption
parameter: $\bar{\gamma }=3.8$ (open circles), $\bar{\gamma }
=5.2$ (full circles) and $\bar{\gamma } =6.7$ (open triangles),
respectively. Each corresponding theoretical distribution $P(v)$
evaluated from Eq.~(5) is also shown: $\gamma =3.8$ (dotted line),
$\gamma =5.2$ (solid line), and $\gamma =6.7$ (dashed line),
respectively.}\label{Fig3}
\end{center}
\end{figure}

Figure~3 shows experimental distributions $P(v)$ for three mean
values of the parameter $\bar{\gamma }$, viz., 3.8, 5.2, and 6.7.
The distribution for $\bar{\gamma }=3.8$ is obtained by averaging
over 69 realizations of microwave graphs having $\gamma $ within
the window $[3.5, \, 4.1]$. The distribution for $\bar{\gamma
}=5.2$ is obtained by averaging over 60 realizations of microwave
graphs having $\gamma $ within the window $[4.7, \, 5.6]$. The
distribution for $\bar{\gamma }=6.7$ is obtained by averaging over
55 realizations of microwave graphs having $\gamma $ within the
window $[6.3, \, 7.1]$. We estimated the experimental values of
the $\gamma$ parameter by adjusting the theoretical mean
reflection coefficient $\langle R \rangle _{th}$ to the
experimental one $\langle R_0 \rangle=\langle S_0S_0^{\dag}\rangle
$, where
\begin{equation}
\label{Eq.7} \langle R \rangle _{th} = \int _0^1dRRP(R).
\end{equation}
We also applied the following interpolation formula
\cite{Kuhl2005} for the distribution $P(R)$:
\begin{equation}
\label{Eq.8} P(R) =
N_{\beta}\frac{e^{-\frac{\alpha}{1-R}}}{(1-R)^{2+\beta /2}}[
A\alpha^{\beta /2 } +B(1-R)^{\beta/2}].
\end{equation}

We offer the following comment on the validity of the Eq.~(8). We
used it instead of exact formulas (12-14) recently presented in
\cite{Savin2005}, which may be used to find the distribution
$P(R)$, because Eq.~(8) is sufficiently accurate (see Fig.~1 in
\cite{Savin2005}) while allowing for much faster numerical
calculations.

Figure~3 also presents for comparison with each experimental
distribution $P(v)$ (symbols) the corresponding numerical distribution
(lines) evaluated  from Eq.~(5).  We see that the experimental
distribution $P(v)$ at $\bar{\gamma } =3.8$ and at 5.2 agree well
with their theoretical counterparts.  However, the comparison for
$\bar{\gamma } =6.7$ shows some discrepancies, particularly
in the range  $0.3<v<0.8$.

\begin{figure}[!]
\begin{center}
\rotatebox{270} {\includegraphics[width=0.5\textwidth,
height=0.6\textheight, keepaspectratio]{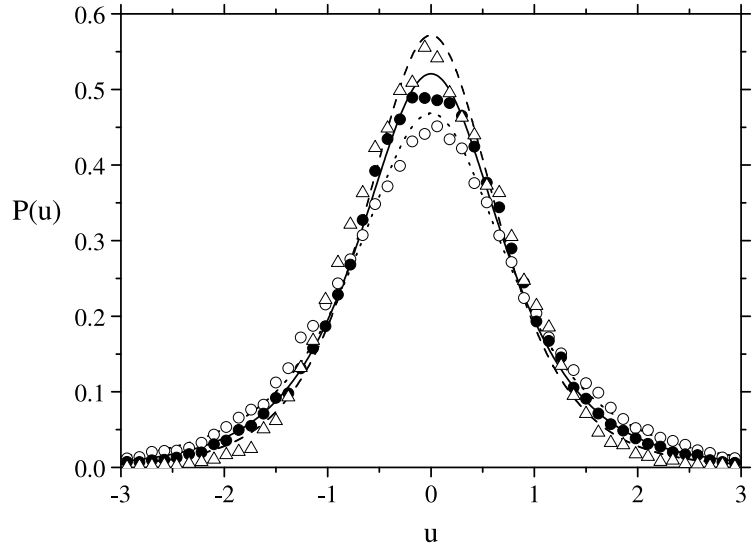}}
\caption{Experimental distribution $P(u)$ of the real part of the
$K$ matrix  at different values of the mean absorption parameter:
$\bar{\gamma }=3.8$ (open circles),  $\bar{\gamma } =5.2$ (full
circles) and $\bar{\gamma }=6.7$ (open triangles), respectively.
Each corresponding theoretical distribution $P(u)$ evaluated from
Eq.~(6) is also shown: $\gamma =3.8$ (dotted line), $\gamma =5.2$
(solid line), and $\gamma =6.7$ (dashed line),
respectively.}\label{Fig4}
\end{center}
\end{figure}

We may use measurements of the distribution $P(u)$ of the real
part of Wigner's reaction  matrix for an imporant and natural
consistency check on our determination of $\gamma$. Figure~4
compares experimental and theoretical $P(u)$ distributions at the
aforementioned values of $\bar{\gamma }$, viz., 3.8, 5.2, and 6.7.
Though each case shows good overall agreement between experimental
and theoretical results, for all three cases the middle
($-0.25<u<0.25$) of the theoretical distribution is slightly
higher than its experimental counterpart.  According to the
definition of the $K$ matrix (see Eq.~(1)), such behavior of the
experimental distribution $P(u)$ suggests a deficit of small
values of $|\textrm{Im} \, S_0|$.  We do not yet know the origin
of this deficit.

Though there are the small discrepancies we have mentioned, the
good overall agreement between experimental and theoretical
results justifies \textit{a posteriori} the procedure we have used
to determine the experimental values of $\gamma$.

The distributions $P(v)$ and $P(u)$ of imaginary and real parts of
Wigner's reaction matrix may be also found using the alternative
approach described in \cite{Anlage2005,Anlage2005b}. In these
papers  the radiation impedance approach was developed and used to
obtaining the distributions of real and imaginary parts of the
normalized impedance
\begin{equation}
\label{Eq.9} Z = \frac{\textrm{Re } Z_{c}+i(\textrm{Im }
Z_{c}-\textrm{Im } Z_{r})}{\textrm{Re } Z_{r}}
\end{equation}
of a chaotic microwave cavity, where
$Z_{c(r)}=Z_0(1+S_{c(r)})/(1-S_{c(r)})$ is the cavity (radiation)
impedance expressed by the cavity (radiation) scattering matrix
$S_{c(r)}$ and $Z_0$ is the characteristic impedance of the
transmission line. The radiation impedance $Z_{r}$ is the
impedance seen at the input of the coupling structure for the same
coupling geometry, but with the sidewalls removed to infinity.
This interesting approach is especially useful in the studies of
microwave systems, in which, in general, both the system and
radiation impedances  are measurable. However, it is not obvious
how to use in practice this approach in the case of quantum
systems.

We used this alternative approach to find distributions $P(v)$ and
$P(u)$ of imaginary and real parts of Wigner's reaction  matrix
for irregular tetrahedral microwave graphs. Wigner's reaction
matrix can be simply expressed by the normalized impedance
$K=-iZ$. The radiation impedance $Z_{r}$ was found experimentally
by measuring in two different frequency windows, viz., 3.5--7.5
GHz and 12-16 GHz of the scattering matrix $S_{r}$ of the 4-joint
connector with three joints terminated by 50 $\Omega $
terminators.

\begin{figure}[!]
\begin{center}
\rotatebox{270} {\includegraphics[width=0.5\textwidth,
height=0.6\textheight, keepaspectratio]{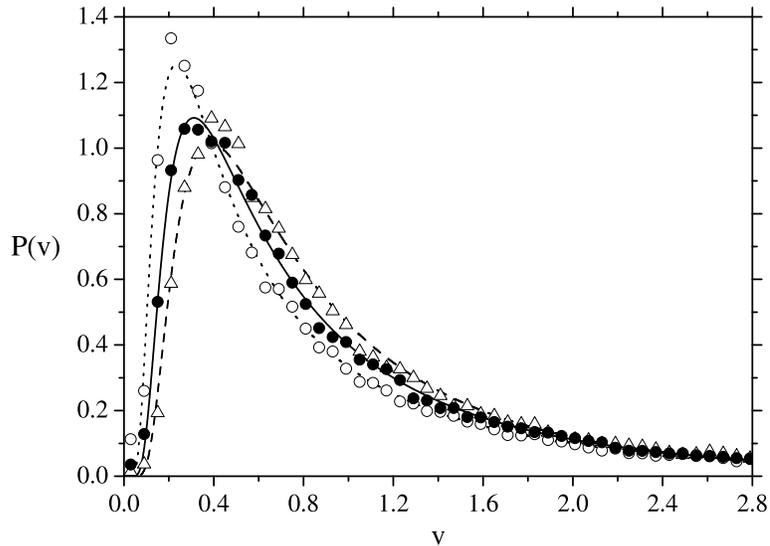}}
\caption{Experimental distribution $P(v)$ of the imaginary part of
the $K$ matrix  at different values of the mean absorption
parameter: $\bar{\gamma }=3.8$ (open circles), $\bar{\gamma }
=5.2$ (full circles) and $\bar{\gamma } =6.7$ (open triangles),
respectively, calculated using the radiation impedance approach
\cite{Anlage2005,Anlage2005b}. Each corresponding theoretical
distribution $P(v)$ evaluated from Eq.~(5) is also shown: $\gamma
=3.8$ (dotted line), $\gamma =5.2$ (solid line), and $\gamma =6.7$
(dashed line), respectively.}\label{Fig5}
\end{center}
\end{figure}

\begin{figure}[!]
\begin{center}
\rotatebox{270} {\includegraphics[width=0.5\textwidth,
height=0.6\textheight, keepaspectratio]{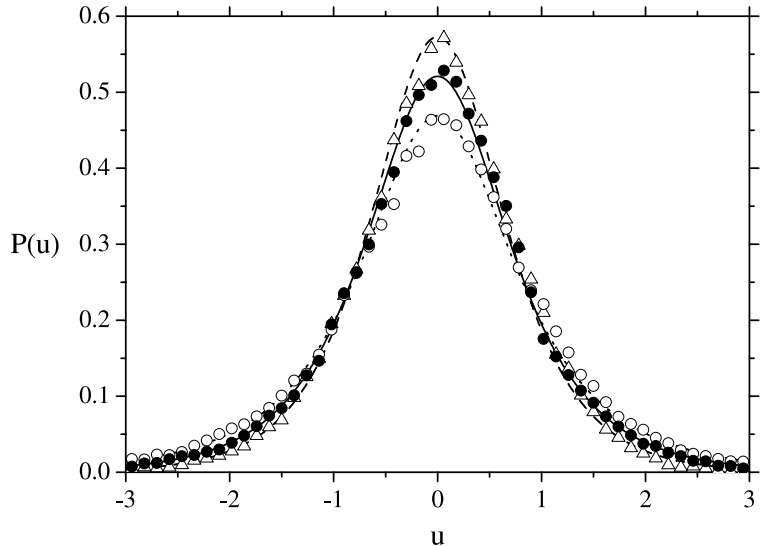}}
\caption{Experimental distribution $P(u)$ of the real part of the
$K$ matrix  at different values of the mean absorption parameter:
$\bar{\gamma }=3.8$ (open circles), $\bar{\gamma } =5.2$ (full
circles) and $\bar{\gamma }=6.7$ (open triangles), respectively,
calculated using the radiation impedance approach
\cite{Anlage2005,Anlage2005b}. Each corresponding theoretical
distribution $P(u)$ evaluated from Eq.~(6) is also shown: $\gamma
=3.8$ (dotted line), $\gamma =5.2$ (solid line), and $\gamma =6.7$
(dashed line), respectively.}\label{Fig6}
\end{center}
\end{figure}

In Fig.~5 and Fig.~6 we show the distributions $P(v)$ and $P(u)$
calculated using the radiation impedance approach
\cite{Anlage2005,Anlage2005b}. As in  the case of the scattering
matrix approach, the experimental distributions are obtained at
three values of the parameter $\bar{\gamma } =3.8$, 5.2 and 6.7.
Figure~5 shows that  the distribution $P(v)$ of the imaginary part
of Wigner's reaction matrix for $\bar{\gamma } = 5.2$ is in good
agreement with the theoretical prediction \cite{Fyodorov2004}.
However, for $\bar{\gamma } =3.8$ and 6.7 the theoretical results
are slightly higher than the experimental ones, what is especially
noticeable at the peaks of the distributions. The experimental
distribution $P(u)$ of the real part of Wigner's reaction matrix
presented in Figure~6 at three values of the parameter
$\bar{\gamma }=3.8$, 5.2 and 6.7 displays a very good agreement
with the theoretical result. The comparison of Figures 3 and 5 and
Figures 4 and 6 show that the distributions $P(v)$ and $P(u)$
evaluated by means of the radiation impedance approach are at the
peaks slightly higher than the ones obtained by the scattering
matrix approach, what may suggest that the influence of the phase
of $S_{r}$ on the distributions is not negligible
\cite{Anlage2005b}.

In summary, using the scattering matrix approach and the radiation
impedance approach we have measured distributions $P(v)$ and
$P(u)$ of imaginary and real parts of Wigner's reaction matrix for
irregular tetrahedral microwave graphs consisting of SMA cables,
connectors, and attenuators.  Use of different attenuators allowed
us to vary absorption in the graphs in a controlled, quantitative
way. For the case of time reversal symmetry ($\beta=1$), the
experimental results for $P(v)$ and $P(u)$ calculated for both
approaches at the same three values of the mean parameter
$\bar{\gamma }$ are in good overall agreement with theoretical
predictions.

{\bf Acknowledgments:} This work was supported by KBN grant No. 2
P03B 047 24 and an equipment grant from ONR(DURIP).

\end{document}